\documentclass[journal]{IEEEtran}
\usepackage{amsmath,amssymb}
\usepackage{amsfonts}
\usepackage{amsbsy}
\usepackage{graphicx}

\usepackage{subfigure}
\usepackage{stfloats}
\usepackage{epstopdf}
\usepackage{color}
\usepackage{etoolbox}
\makeatletter
\patchcmd{\@makecaption}
{\scshape}
{}
{}
{}
\makeatother
\def\tablename{Table}
\begin{document}

\title{Coverage Probability of 3D Mobile UAV Networks}

\author{\IEEEauthorblockN{Pankaj K. Sharma and Dong In Kim} 
\thanks{The authors are with the College of Information and Communication Engineering, Sungkyunkwan University (SKKU), Korea. Email:  \{sharmapk,$\,$dikim\}@ieee.org.}
%\thanks{This work is supported by the National Research Foundation of Korea (NRF) Grant funded by the Korean Government under Grant 2017R1A2B2003953.}
}
\maketitle

\begin{abstract}
In this paper, we consider a network of multiple unmanned aerial vehicles (UAVs) where a given number of UAVs are placed at three-dimensional (3D) locations in a finite circular disk shaped region to serve a reference ground user equipment (UE) located at its center. 
Herein, a serving UAV is assumed to be located at fixed altitude which communicates with the reference UE. All the other UAVs in the network are designated as interfering UAVs to the UE and are assumed to have 3D mobility. 
To characterize the 3D UAV movement process, we hereby propose an effective 3D mobility model based on the mixed random waypoint mobility (RWPM) and uniform mobility (UM) models in the vertical and spatial directions. 
Further, considering the proposed 3D mobility model, we first characterize the interference received at reference UE, and then evaluate its coverage probability under Nakagami-\emph{m} fading.  
We quantify the achievable performance gains for the ground UE under various system and channel conditions. Moreover, we corroborate our analytical results through simulations.

\end{abstract}

\section{Introduction}
Unmanned aerial vehicles (UAVs) have emerged as key enabler for seamless wireless connectivity in diverse scenarios such as grand temporary events, military operations and disaster situations \cite{zeng}, and are anticipated to be potential candidates for future fifth-generation ($5$G) networks. Due to miniaturized structure and low-cost deployment, the UAVs are considered as flying wireless access platforms in three-dimensional (3D) space.
%These UAV platforms are broadly classified as high altitude platforms (HAP) and low altitude platforms (LAP) based on their size and operating range. While the HAP has an operating altitude range of hundreds of kilometers, the LAP operates at altitude range of few hundred meters.
Of particular interest are the rotary wing type UAVs which can precisely hover over a 3D location in space. The UAVs possess high possibility of line-of-sight (LoS) connection with a ground user equipment (UE) \cite{kan}. Due to mobility, the UAVs can flexibly control their altitude and spatial location to enhance UE's quality of service (QoS) requirements \cite{han}. 
%Among various applications, the deployment of multiple UAVs for providing ubiquitous wireless coverage in cellular networks is of great importance, especially where conventional infrastructure is not accessible \cite{zeng} e.g., remote areas, temporary hotspot areas, etc. In general, a multiple UAV network comprised of a given number of UAVs placed at 3D locations in a geographical region to serve ground UEs. 
Unlike static networks, the UAV nodes dynamically adjust their altitude and trajectory parameters in order to enhance the system performance. Consequently, UAV networks are time varying in nature where the ground UEs receive fluctuating interference from interfering UAVs. Since UAVs possess 3D mobility, the interference patterns at ground UEs cannot be effectively characterized by the existing results for two-dimensional (2D) terrestrial cellular networks. Thus, the analysis of coverage probability of 3D mobility-based UAV networks is an open issue.

Towards this end, in this letter, we first propose a 3D mobility model (to be described later) for UAV movement process based on the mixed random waypoint mobility (RWPM) \cite{bet} and uniform mobility (UM) \cite{zhen} models. Then, by following the proposed 3D mobility model, we characterize the interference received at a reference ground UE and evaluate its coverage probability under Nakagami-\emph{m} fading. To our best knowledge, we are the first to apply the 3D mobility based approach to analyze the coverage performance of a network of multiple UAVs being deployed.
%The RWP mobility is a well-accepted model for mobile ad hoc network analysis which is quite attractive for modelling UAV's vertical movement process.

%\subsection{Related Work}
Recently, 
%UAV networks have gained significant research interests and are actively investigated in the literature. For instance, 
the deployment of single UAV has been investigated in \cite{kan} and multiple UAVs in \cite{moz}-\cite{hlp} for coverage enhancement.
%The work in \cite{ruii} has deployed a single UAV as flying base station to serve multiple ground users using cyclical multiple access scheme. In \cite{sje}, the authors have analyzed joint uplink and downlink communications via UAV mounted cloudlet.
%The deployment of multiple UAVs for coverage extension is addressed in \cite{moz}-\cite{hlp}.
Specifically, the authors in \cite{moz} have considered rotary wing type UAVs while the authors in \cite{rui} have considered fixed wing type UAVs at constant altitude. In \cite{hlp}, the authors have analyzed the 3D deployment of UAVs 
%using geometric disk cover approach 
in an interference-free environment. Further, the works in \cite{moz2}, \cite{czh} have analyzed spectrum sharing in UAV networks. {The work in \cite{lyu2} has investigated the offloading for cellular hotspot in spectrum sharing UAV networks.} In contrast, several works have focused on the performance analysis of UAV networks. The authors in \cite{vv} have analyzed the coverage performance of a finite 3D UAV network. However, the multiple UAVs are deployed in a disk region at a constant height. In \cite{bgal}, the coverage performance of a multiple UAV network has been analyzed with wireless backhaul. In \cite{mmaz}, the coexistence of terrestrial users and aerial UAV UE in cellular networks has been investigated. The authors in \cite{rjly} have evaluated the blocking probability of a UAV in a network of multiple UAVs.

{Note that the aforementioned works have relied on the modeling of the UAV locations as point process (e.g., homogeneous Poisson point process (PPP) or binomial point process (BPP)). However, such approach may not sufficiently capture the realistic 3D deployment of UAVs, especially if both the spatial movement and the altitude control mechanism do not necessarily follow the same random process.} 
%Further, the analysis of such case for finite 3D UAV networks are mathematically challenging.} 
None of these works have considered 3D mobility of UAVs to analyze the coverage probability of a reference UE in UAV networks.

%Motivated by this, in this work, we propose an effective 3D mobility-based modelling for UAV movement process to characterize the resulting interference and coverage probability at a reference ground UE.
%Here, the instantaneous altitude of the interfering UAVs is determined using steady state non-uniform distribution of the RWP mobility process.

%Unlike aforementioned works, the UAV has been employed as a mobile relay in \cite{rui2}, \cite{hdk} to assist the communications between two distant ground users. .

%\textit{Notations}: $\mathbb{E}\{\cdot\}$ denotes expectation and $\mathcal{CN}(\mu,\sigma^{2})$ denotes complex normal distribution with mean $\mu$ and variance $\sigma^{2}$.

\section{System, Mobility and Channel Models}\label{sysmod}
\setlength{\belowdisplayskip}{3pt}
\setlength{\abovedisplayskip}{3pt}
\subsection{System Model}
We consider a 3D UAV network with $M+1$ UAVs $S_{i}$, $i\in\{0,...,M\}$, where a UAV $S_0$ is a serving UAV to a reference UE $U$ at stationary altitude
%\footnote{Herein, as our focus is to propose a 3D mobility-based modeling for the performance analysis of UAV networks, we ignore the case of random UAV selection. We defer the analysis for such case to future works.}
$h_0$ above the ground. We assume that both the 2D projection of $S_0$ and the location of $U$ are at origin. 
%\footnote{\textcolor{red}{Here, we ignore the random serving UAV distance which could be a possible extension for future works.}
All the remaining $M$ UAVs act as the interfering UAVs to $U$ and are initially launched uniform at random in a 3D cylindrical space with radius $R$ and height $H$ with $\{z_i\}^{M}_{i=1}\equiv\Phi\subset\mathbb{R}^{2}$ as their projections in 2D plane.
%All the remaining $M$ UAVs act as the interfering UAVs to $U$ and are initially launched uniform at random such that their projections $\{z_i\}^{M}_{i=1}\equiv\Phi\subset\mathbb{R}^{2}$ in a 2D circular region of radius $R$.
%The locations of interfering UAVs along this circular periphery is determined uniformly at random in initial time slot. We also assume that after initial deployment the UAVs cannot change their horizontal locations for entire observation time.
%Considering that the directional antenna at each UAV has directional antenna pointing towards ground with half-power beamwidths of $2\phi$. Consequently, a UAV $S_m$ illuminates a disc shaped coverage area on the ground with radius $z_m=h_m\tan\phi$. Note that the radius $z_m$ of the coverage area beneath UAV $S_m$ depends upon its current height. Hence, for UAV $S_m$, if $z_m>R$, it interferes with the $U$.
As UAVs dynamically control their altitude and spatial locations, a 3D mobility model for UAV movement is of prime importance.  Therefore, we propose a mixed mobility (MM) model to characterize 3D UAV movement process in vertical and spatial directions based on the well-known RWPM and UM \cite{zhen} models, respectively.
\begin{figure}[!t]
\centering
\includegraphics[width=3in]{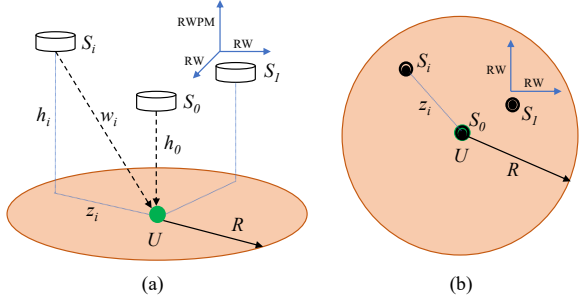}
\caption{(a) Multiple UAVs network with 3D mobility. (b) Projection of the UAVs locations on 2D plane.}
\label{sys1}
\end{figure}
%\begin{figure}[!t]
%\centering
%\includegraphics[width=2.3in]{syst2.pdf}
%\caption{Projection of 3D position and the coverage area of UAVs in 2D plane.}
%\label{sys2}
%\end{figure}

\subsection{Mixed Mobility Model for 3D UAV Movement Process}
In MM model, at any time $t$, each interfering UAV either adjusts its altitude $h_i\in[0,H]$ following RWPM or make a spatial excursion in the circular region following UM.
%Specifically, we consider the RWPM model with think time for UAVs' vertical movement process.
%Note that, in MM model, the UAV location transition is allowed only in one direction at a time. This makes our MM model simple yet effective for characterizing the interference and evaluating the coverage probability of a 3D UAV network.
Specifically, we consider the RWPM model with dwell time \cite{bet} for the vertical movement of UAVs in a finite interval $[0,H]$. Initially, the UAV $S_i$ is placed at height $H_1$ selected uniformly random from the interval $[0,H]$. Then, this UAV located at $H_1$ selects a random waypoint at $H_2$ uniformly random in $[0,H]$ and move towards it with constant velocity $v_{1,i}$ chosen uniformly random from $[v_{min},v_{max}]$. Once the UAV reaches at waypoint $H_2$, it may stay here for a dwell time $T_\mathrm{s}$ drawn from a uniform distribution $[\tau_{min}, \tau_{max}]$ and then, repeats again the same procedure to find next waypoint and so on. With $p_\mathrm{s}$ as staying probability, at time instant $t$ in steady state, the static probability density function (pdf) of altitude $h_i$ is given by $p_\mathrm{s}f^{st}_{h_i}(x|t)$  and the non-uniform mobility pdf is given by $(1-p_\mathrm{s})f^{mo}_{h_i}(x|t)$,
%\begin{align}\label{1}
%f_{h_{i}}(x)&=p_\mathrm{s}f^{st}_{h_i}(x)+(1-p_\mathrm{s})f^{mo}_{h_i}(x),
%\end{align}
where $f^{st}_{h_{i}}(x|t)=\frac{1}{H}$ and
\begin{align}\label{1y}
f^{mo}_{h_{i}}(x|t)&=-\frac{6x^2}{H^3}+\frac{6x}{H^2}, \textmd{ for } 0 \leq x \leq H
\end{align}
with
%\begin{align}\label{rrwp}
$p_\mathrm{s}=\frac{\mathbb{E}[T_\mathrm{s}]}{\mathbb{E}[T_\mathrm{s}]+\mathbb{E}[T_\mathrm{m}]}$,
%\end{align}
$\mathbb{E}[T_\mathrm{s}]$ is the mean stay time, $\mathbb{E}[T_\mathrm{m}]=\mathbb{E}[\frac{L}{v_{1,i}}]=\frac{\ln(v_{max}/v_{min})}{v_{max}-v_{min}}\mathbb{E}[L]$ \cite{bet} is the mean movement time between two consecutive staying periods, and $\mathbb{E}[L]=\frac{H}{3}$ is the mean RWPM leg length.

Further, in MM model, the UAVs make spatial movement utilizing this dwell time period $T_\mathrm{s}$ at a waypoint. Hereby, we consider random walk (RW) as UM model for the spatial excursions. With RW model, at any time instant $t$, a UAV independently and randomly selects a new direction and start moving with random speed. Thus, at time instant $t$ during dwell time period, the UAV $S_i$ makes transition to new spatial location $z_i$ with probability $p_\mathrm{s}$ according to $z_i(t+1)=z_{i}(t)+u_i(t)$, where $u_i(t)$ is uniformly at random in ball $B(z_i(t),R^\prime)$ with $R^\prime$ as the mobility range. While, it stays at previous spatial location with probability $1-p_\mathrm{s}$ as $z_i(t+1)=z_i(t)$. Let ${v}_{2,i}=\mathbb{E}[\|z_i(t+1)-z_{i}(t)\|]=\frac{R^\prime}{1.5}$ \cite{zhen} which gives the average speed of spatial movement of UAVs.
%The spatial location of UAV $S_i$ at time $t+1$ is given by
%\begin{align}
%z_i(t+1)=z_i(t)+\bar{v} u_i(t),
%\end{align}
%where $u_i(t)$ is uniformly at random in ball $B(z_i(t),\bar{v}R_{rw})$ with $R_{rw}$ as normalized mobility range and
%\end{align}
%is utilized by within a radius of $a_0$ around initial locations $\{y_i\}$ given a waypoint height .
%Further, in MM model, we assume that this ``dwell time" is utilized by the UAVs for spatial excursion within a radius of $a_0$ around initial locations $\{y_i\}$ given a waypoint height. With center at initial location, the location of a UAV under UM follows a uniform distribution in a circle of radius $a_0$ as
%\begin{align}\label{th11}
%f_{u_i}(x)&=\left\{ \begin{array}{l}
%\frac{1}{\pi a^2_0}, \textmd{ for } \|x\| \leq a_0, \\
%0, \textmd{ else},
%\end{array}\right.
%\end{align}
%where $\|\cdot\|$ denotes the Euclidean distance. Thus, at time $t$, the UAV $S_i$ makes spatial excursion to new spatial location $z_i$ with probability $p_p$ as $z_i(t)=y_{i}+u_i(t)$. While, it stays at previous spatial location with probability $1-p_p$ as $z_i(t)=z_i(t-1)$. Let $v_{2,i}(t)\triangleq\|z_i(t)-z_{i}(t-1)\|$ as velocity of spatial movement of UAV $S_i$, we have $\bar{v}_{2,i}\triangleq\mathbb{E}[v_{2,i}(t)]=128a_{0}/45\pi$, $\forall t$ \cite{zhen}.
%\begin{figure}[!t]
%\centering
%\includegraphics[width=1.5in]{rwp.pdf}
%\caption{Illustration of RWP mobility model.}
%\label{rwp}
%\end{figure}
\subsection{Channel Model and Interference}
At time $t$, the free-space path loss from UAV $S_i$ to $U$ is given by (see \cite{rjly} and reference therein) $w^{-\alpha}_i(t)=\left({h^2_i(t)+Z^2_i(t)}\right)^{-\frac{\alpha}{2}}$,
where $w_i$ is the UAV $S_i$ to $U$ distance, $Z_i(t)=\|z_i(t)\|$ is the Euclidean norm, and $\alpha$ is the path loss exponent. Consequently, the aggregate interference at $U$ is given as $I(t)\equiv\sum_{i\in{\Phi_{st}(t)}}g_i(t)w^{-\alpha}_i(t)+\sum_{i\in{\Phi_{mo}(t)}}g_i(t)w^{-\alpha}_i(t)$, $|\Phi_{st}(t)\cup\Phi_{mo}(t)|=M$ where $\Phi_{st}(t)$ and $\Phi_{mo}(t)$ are the sets containing a number of UAVs which at time $t$ made  the spatial movement and the vertical movement, respectively. The resulting signal-to-interference-ratio (SIR) at $U$ can be given as $\Lambda(t)=\frac{g_0(t) h^{-\alpha}_{0}}{I(t)}$,
where $g_0(t)$ and $g_i(t)$ are the fading channel coefficients from $S_0$ and $S_i$ to $U$, respectively. Further, for Nakagami-\emph{m} fading, the channel gain $g_i$ follows a gamma distribution whose pdf is given by
\begin{align}
f_{g_i}(x)&=\frac{m^{m_i}_ix^{m_i-1}}{\Gamma(m_i)}\textmd{exp}(-m_i x),
\end{align}
where $m_i$ represents an integer-valued fading parameter. Hereafter, we drop the time notation $t$ from $I(t)$ and $\Lambda(t)$ since our focus is the coverage probability analysis in a single snapshot.
\section{Coverage Probability Analysis}
For a target SIR threshold $\psi$, the coverage probability $P_{cov}$ at $U$ is given by
\begin{align}\label{pco}
P_{cov}&=\textmd{Pr}[\Lambda>\psi].
\end{align}
\newtheorem{theorem}{Theorem}
\begin{theorem}\label{th3}
For Nakagami-\emph{m} fading, the coverage probability $P_{cov}$ in (\ref{pco}) can be expressed as
\begin{align}\label{pcoc}
P_{cov}&=\sum_{k=0}^{m_0-1}\frac{(-m_0\psi h^\alpha_0)^k}{k!}\left[\frac{\partial^k}{\partial s^k}\mathcal{L}_I(s)\right]_{s=m_0\psi h^{\alpha}_0},
\end{align}
where $\mathcal{L}_I(s)$ is the Laplace transform of the interference power which is given by
	\begin{align}\label{lt}
	&\mathcal{L}_I(s)=\sum_{n=0}^{M}\binom{M}{n}[p_\mathrm{s}{\Upsilon_{st}(s)}]^{n}[(1-p_\mathrm{s})\Upsilon_{mo}(s)]^{M-n},
	\end{align}
	with $\Upsilon_{st}(s)$ and $\Upsilon_{mo}(s)$ as
	\begin{align}\label{stt}
	\Upsilon_{st}(s)&\triangleq{\mathbb{E}_{W_i}\bigg(\frac{m_i}{m_i+sw^{-\alpha}_i}\bigg)^{m_i}}, i\in\Phi_{st},
	\end{align}
   \begin{align}\label{mot}
	\Upsilon_{mo}(s)&\triangleq{\mathbb{E}_{W_j}\bigg(\frac{m_j}{m_j+sw^{-\alpha}_j}\bigg)^{m_j}}, j\in\Phi_{mo}.
	\end{align}
\end{theorem}
\begin{IEEEproof}
	See Appendix \ref{AA}.
\end{IEEEproof}

According to Theorem \ref{th3}, to evaluate $P_{cov}$, we need to derive the analytical expressions of $\Upsilon_{st}(s)$ and $\Upsilon_{mo}(s)$. 
To proceed further, we first determine the cumulative distribution function (cdf) and pdf of distance $W_i$ for $i\in\Phi_{st}$ or $\Phi_{mo}$ in the sequel.
\newtheorem{lemma}{Lemma}
\newtheorem{proposition}{Proposition}
\begin{proposition}\label{tho1}
The cdf of the distance $W_i$ from $S_i$ to $U$ when $S_i$ makes spatial excursion i.e., $i\in\Phi_{st}$ is given by
\begin{align}\label{th101}
F^{st}_{W_i}(w_i)&=\left\{ \begin{array}{l}
\frac{2}{3}\frac{w^{3}_i}{R^2 H},\textmd{ for }0\leq w_i<H,\\
\frac{w^2_i}{R^2}-\frac{1}{3}\frac{H^2}{R^2},\textmd{ for }H\leq w_i<R,\\
\frac{w^2_i}{R^2}-\frac{1}{3}\frac{H^2}{R^2}-\frac{2}{3}\frac{(w^2_i-R^2)^{\frac{3}{2}}}{R^2 H},\\
\qquad \quad\textmd{ for } R\leq w_i\leq\sqrt{R^2+H^2}.
\end{array}\right.
\end{align}
\end{proposition}
\begin{IEEEproof}
See Appendix \ref{appA}.
\end{IEEEproof}
\newtheorem{corollary}{Corollary}
\begin{corollary}\label{jh11}
	The corresponding pdf of the distance $W_i$ if $S_i$ makes spatial excursion can be expressed as  
	\begin{align}\label{j11}
f^{st}_{W_i}(w_i)&=\left\{ \begin{array}{l}
\frac{2w^{2}_i}{R^2 H},\textmd{ for }0\leq w_i<H,\\
\frac{2w_i}{R^2},\textmd{ for }H\leq w_i<R,\\
\frac{2w_i}{R^2}-\frac{2w_i\sqrt{w^2_i-R^2}}{R^2H},\\
\qquad\textmd{ for } R\leq w_i\leq\sqrt{R^2+H^2} .
\end{array}\right.
\end{align}
\end{corollary}
\begin{IEEEproof}
	Taking the derivative of $F^{st}_{W_i}(w_i)$ in (\ref{th101}) with respect to $w_i$ yields the pdf $f^{st}_{W_i}(w_i)$ in (\ref{j11}).
\end{IEEEproof}
\begin{proposition}\label{th2}
	The cdf of the distance $W_i$ from $S_i$ to $U$ if $S_i$ makes vertical movement i.e., $i\in\Phi_{mo}$ is given by
\begin{align}\label{wmn}
F^{mo}_{W_i}(w_i)&=\!\!\left\{ \begin{array}{l}
-\frac{4}{5}\frac{w^{5}_i}{R^2 H^3}+\frac{3}{2}\frac{w^{4}_i}{R^2 H^2},\textmd{ for }0\leq w_i<H,\\
\frac{w^2_i}{R^2}-\frac{3}{10}\frac{H^2}{R^2},\textmd{ for }H\leq w_i<R,\\
\frac{w^2_i}{R^2}\!-\!\frac{3}{10}\frac{H^2}{R^2}\!-\!\frac{3}{2}\frac{(w^2_i-R^2)^2}{R^2 H^2}\!+\!\frac{4}{5}\frac{(w^2_i-R^2)^{\frac{5}{2}}}{R^2 H^3}\\
\qquad\textmd{ for } R\leq w_i\leq\sqrt{R^2+H^2}.
%0, \textmd{ else}.
\end{array}\right.
\end{align}
\end{proposition}
\begin{IEEEproof}
The derivation of Lemma \ref{th2} is similar to that in Appendix \ref{appA} with the pdf $f^{st}_{h_i}(x)$ replaced by $f^{mo}_{h_i}(x)$.
	\end{IEEEproof}
\begin{corollary}
	The corresponding pdf of the distance $W_i$ if $S_i$ undergo vertical movement is given by  
	\begin{align}\label{tho11}
	f^{mo}_{W_i}(w_i)&=\left\{ \begin{array}{l}
	-\frac{4w^{4}_i}{R^2 H^3}+\frac{6w^{3}_i}{R^2 H^2},\textmd{ for }0\leq w_i<H,\\
	\frac{2w_i}{R^2},\textmd{ for }H\leq w_i<R,\\
	\frac{2w_i}{R^2}-\frac{6w_i(w^2_i-R^2)}{R^2 H^2}+\frac{4w_i(w^2_i-R^2)^{\frac{3}{2}}}{R^2 H^3},\\
	\qquad \quad\textmd{ for } R\leq w_i\leq\sqrt{R^2+H^2}.
	%0, \textmd{ else}.
	\end{array}\right.
	\end{align}
\end{corollary}
\begin{IEEEproof}
	The proof is straightforward by taking the derivative of cdf $F^{mo}_{W_i}(w_i)$ in (\ref{wmn}) with respect to $w_i$.
\end{IEEEproof}

Having derived the relevant UAV distance distributions, we now proceed to evaluate the term $\Upsilon_{st}(s)$ in (\ref{lt}). Accordingly, we invoke the pdf $f^{st}_{W_i}(w_i)$ from (\ref{j11}) into (\ref{stt}) which yields
\begin{align}\label{fboh}
&\Upsilon_{st}(s)=\sum_{l=0}^{m_i}\binom{m_i}{l}(-1)^l[\mathcal{I}_{l}(a_1,a_2,\ell_1,3)\\\nonumber
&+\mathcal{I}_{l}(a_2,a_3,2,2)+\mathcal{I}_{l}(a_3,a_4,2,2)-\mathcal{J}_{l}(a_3,a_4,\ell_2,1)],
\end{align}
where the functions $\mathcal{I}_{l}(a,b,\ell,\kappa)$ and $\mathcal{J}_{l}(a_3,a_4,\ell,\kappa)$ are obtained (after change of variable $y=w^{\alpha}_i$), respectively, as
\begin{align}\label{i1}
\mathcal{I}_{l}(a,b,\ell,\kappa)&=\frac{\ell}{\alpha R^2}\int_{a}^{b}\frac{{y^{\frac{\kappa}{\alpha}-1}}}{(1+m_i s^{-1}y)^l}dy
\end{align}
and
\begin{align}\label{i2}
\mathcal{J}_{l}(a_3,a_4,\ell,\kappa)&=\frac{\ell}{\alpha R^2}\int_{a_3}^{a_4}\frac{y^{\frac{2}{\alpha}-1}{(y^{\frac{2}{\alpha}}-R^2)}^{\frac{\kappa}{2}}}{(1+m_i s^{-1}y)^l}dy
%\end{align}
\end{align}
with $a_1=0$, $a_2=H^\alpha$, $a_3=R^\alpha$, $a_4=(R^2+H^2)^{\alpha/2}$, $\ell_1=\ell_2=\frac{2}{H}$. Here, we consider $\alpha=2$\footnote{The value $\alpha=2$ for path loss exponent is commonly adopted in literature and also observed in an industrial UAV field trial report (refer \cite{rjly} for details).} to obtain the closed-form solutions of the integral expressions in (\ref{i1}) and (\ref{i2}). For $\alpha=2$, (\ref{i1}) can be evaluated as \cite[eq. 1.2.4.3]{pru}
\begin{align}
\mathcal{I}_{l}(a,b,\ell,\kappa)&=\frac{\ell}{R^2}\frac{b^{\frac{\kappa}{2}}}{\kappa}~_2F_1\left(l,\frac{\kappa}{2};1+\frac{\kappa}{2};-m_is^{-1}b\right)\\\nonumber
&-\frac{\ell}{R^2}\frac{a^{\frac{\kappa}{2}}}{\kappa}~_2F_1\left(l,\frac{\kappa}{2};1+\frac{\kappa}{2};-m_is^{-1}a\right),
\end{align}  
where $_{2}F_{1}(\cdot,\cdot;\cdot;\cdot)$ denotes Gauss hypergeometric function \cite{pru}. 
Further, for $\alpha=2$, one can evaluate (\ref{i2}) as
\begin{align}\label{juki}
&\mathcal{J}_{l}(a_3,a_4,\ell,\kappa)=\frac{\ell}{2R^2}\int_{R^2}^{(R^2+H^2)}\frac{({y-R^2})^{\frac{\kappa}{2}}}{(1+m_i s^{-1}y)^l}dy\\\nonumber
&\stackrel{(a)}{=}\frac{\ell}{2R^2}\int_{0}^{H^2}\frac{(m^{-1}_i s)^l{y}^{\frac{\kappa}{2}}}{(R^2+m^{-1}_i s+y)^l}dy\\\nonumber
&\stackrel{(b)}{=}\frac{\ell(\kappa+2)^{-1} H^{\kappa+2}}{R^2(1+m_i s^{-1}R^2)^l}~_{2}F_{1}\left(l,\frac{\kappa}{2}+1;\frac{\kappa}{2}+2;\frac{-H^2}{R^2+m^{-1}_i s}\right),
\end{align}
where in $(a)$, the variable is changed as $z=y-R^2$ and then, $(b)$ follows \cite[eq. 1.2.4.3]{pru}.
Next, the term $\Upsilon_2(s)$ can be evaluated by using the pdf $f^{mo}_{W_i}(s)$ from (\ref{tho11}) into (\ref{mot}) as
\begin{align}\label{mosf}
&\Upsilon_{mo}(s)=\sum_{l=0}^{m_i}\binom{m_i}{l}(-1)^l[\mathcal{I}_{l}(a_1,a_2,\ell_3,4)\\\nonumber
&-\mathcal{I}_{l}(a_1,a_2,\ell_4,5)+\mathcal{I}_{l}(a_2,a_3,2,2)+\mathcal{I}_{l}(a_3,a_4,2,2)\\\nonumber
&-\mathcal{I}_{l}(a_3,a_4,\ell_3,4)+\mathcal{I}_{l}(a_3,a_4,\ell_5,2)+\mathcal{J}_{l}(a_3,a_4,\ell_4,3)],
\end{align}
where the functions $\mathcal{I}_{l}(\cdot,\cdot,\cdot,\cdot)$ and $\mathcal{J}_{l}(\cdot,\cdot,\cdot,\cdot)$ are the same as defined previously with $\ell_3=\frac{6}{H^2}$, $\ell_4=\frac{4}{H^3}$, and $\ell_5=\frac{6R^2}{H^2}$.

Finally, first inserting $\Upsilon_{st}(s)$ and $\Upsilon_{mo}(s)$ from (\ref{fboh}) and (\ref{mosf}) in (\ref{lt}) and then, using the result into (\ref{pcoc}) followed by taking the required derivative, $P_{cov}$ can be computed.  

\section{Numerical Results}\label{num}
For numerical results, we set $H=30$m, $R=40$m, and $\alpha=2$. Further, for simulations, we initially let the network run for $10000$ steps to achieve steady state. We consider $v_{1,i}\sim[v_{min},v_{max}]=[0.2,10]$ m/s, $T_{\mathrm{s}}\sim[\tau_{min},\tau_{max}]=[2,6]$ s corresponding to $p_\mathrm{s}=0.5$ and $v_{2,i}=20/3$ m/s $\forall$ $i$, $i\neq0$.

\begin{figure}[t]
\centering
\includegraphics[width=3.0in]{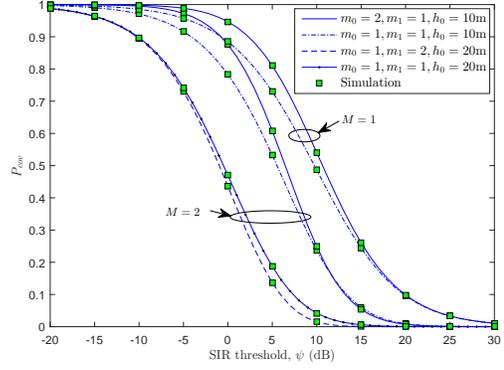}
\caption{{Coverage probability versus SIR threshold for different $M$.}}
\label{pdf}
\end{figure}

\begin{figure}[t]
\centering
\includegraphics[width=3.0in]{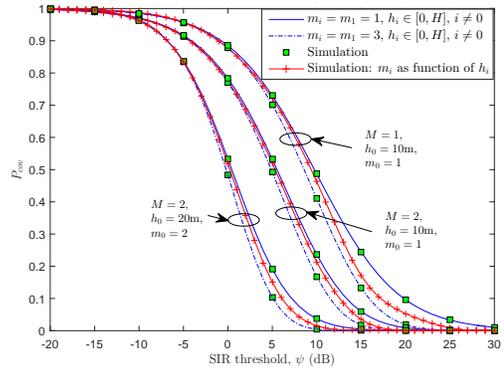}
\caption{{Coverage probability comparison with altitude dependent fading.}}
\label{fi22}
\end{figure}

\begin{table*}[!t]
	\renewcommand{\arraystretch}{1.3}
	\caption{Coverage probability comparison for different values of $p_\mathrm{s}$.}
	\label{tab1}
	\centering
	\begin{tabular}{c||c|c|c|c|c|c}
		\hline\hline
		&$\psi=-20$ dB  &$\psi=-10$ dB  & $\psi=0$ dB &$\psi=10$ dB  &$\psi=20$ dB  &$\psi=30$ dB  \\
		\hline\hline
		$p_\mathrm{s}=0.1$&$0.988266$  &$0.898597$  &$0.468978$  &$0.0413881$  &$0.000674683$  &$7.14876\times10^{-6}$ \\
		\hline 
		$p_\mathrm{s}=0.9$&$0.987466$  &$0.896137$  &$0.471149$  &$0.0426635$  &$0.000703855$ &$7.47065\times10^{-6}$  \\ 
		\hline 
	\end{tabular}
\end{table*}

{Fig. \ref{pdf} plots the coverage probability $P_{cov}$ of reference UE $U$ against different SIR threshold for various system parameters. Here, we set the fading parameter of all interfering UAVs as $m_1$ and the stay probability $p_\mathrm{s}=0.5$. We can observe that the coverage probability degrades when either of $M$ or $m_1$ increases for a given altitude of serving UAV $h_0$. This follows due to increased amount of interference at the UE $U$. Also, for a higher value of $m_0$ (e.g., $m_0=2$) coverage probability improves. Upon increasing the distance $h_0$ from $10$m to $20$m for $M=2,m_1=1$, the coverage probability further degrades.}   

{Fig. \ref{fi22} presents the coverage probability comparison of the considered system with altitude dependent fading conditions. Here, we quantize the fading parameter $m_i$, $i\in\{0,..,M\}$ for the UAVs as $m_i=1$, $2$ and $3$ according to their instantaneous altitude $h_i \in [0,\frac{H}{3}]$, $[\frac{H}{3},\frac{2H}{3}]$ and $[\frac{2H}{3},H]$, respectively. This corresponds to the scenario where LoS path is more likely to occur when the altitude of UAVs takes on a higher value. We have investigated this case through simulations only and the results are compared with our analysis where the parameter $m_i$ is chosen independent of the UAVs' altitude. We can clearly see that our analysis with the values of $m_i$ as $1$ and $3$ for interfering UAVs yields respectively the lower and upper bounds on the coverage probability with altitude dependent fading at given $m_0$.}

{Furthermore, in Table \ref{tab1}, we compare the coverage probability against SIR threshold for different values of $p_\mathrm{s}$. Here, we can observe that the coverage probability with $p_\mathrm{s}=0.9$ is inferior than that with $p_\mathrm{s}=0.1$ at lower values of SIR threshold, but shows marginal improvement as SIR threshold increases beyond certain value.}
%Fig. \ref{fi22} investigates the impact of different values of staying probability $p_\mathrm{s}$ on the coverage probability at UE $U$. Here, we set the number of interferers $M=2$ and the interfering fading channel parameter $m_1=3$ with $m_0=1$. We can see that the coverage probability at $U$ depends upon the value of $p_\mathrm{s}$. For instance, when $p_\mathrm{s}$ increases from $0.1$ to $0.4$ or $0.9$, the coverage probability improves. This reveals that the impact of the $p_\mathrm{s}$ is critical for the coverage performance of UE. In fact, this illustrates that the random spatial excursion for larger time duration is much beneficial for the coverage performance enhancement at $U$ as compared to the vertical movement for the same time duration.    

\section{Conclusion}\label{con}
We have analyzed the coverage probability of a reference UE in a 3D mobile UAV network. Herein, first, we have proposed an effective 3D mobility model by mixing the RWPM and RW for vertical UAV movement and spatial excursions, respectively. Then, based on this mixed mobility (MM) model, we characterized the aggregate interference at a reference UE and derived its coverage probability. We observed that the a higher stay probability at the waypoints is advantageous for coverage performance beyond certain SIR threshold.  %Considering the steady state non-uniform distribution of RWP mobility process and generalized Nakagami-\emph{m} fading, we characterized the node distributions of the distances between interfering UAVs and the ground UE. We derived the coverage probability of the ground UE in single-fold integral form as function of Laplace transform of aggregate interference power. We revealed the impact of various system and channel parameters on the achievable performance gains of the UE.

%\section*{Acknowledgment}
%This works was supported by....
%\section{Acknowledgment}
%This work was supported by the National Research Foundation of Korea (NRF) Grant funded by the Korean Govt. (MSIP) under Grant 2017R1A2B2003953 and LG Electronics, Inc.
\appendices
\section{Proof of Lemma \ref{th3}}\label{AA}%}
As followed in \cite{vv}, the Laplace transform of interference power can be obtained as
\begin{align}\nonumber
&\mathcal{L}_I(s)=\mathbb{E}_I\bigg[\prod_{i\in\Phi_{st}}\textmd{exp}\left(-sg_i w^{-\alpha}_i \right)\prod_{j\in\Phi_{mo}}\textmd{exp}\left(-sg_j w^{-\alpha}_j \right)\bigg]\\\nonumber
&\stackrel{(a)}{=}\mathbb{E}_{W}\mathbb{E}_{g}\bigg[\prod_{i\in\Phi_{st}}\textmd{exp}\left(-sg_i w^{-\alpha}_i \right)\prod_{j\in\Phi_{mo}}\textmd{exp}\left(-sg_j w^{-\alpha}_j \right)\bigg]\\\nonumber
%%\bigg[\prod_{i=1}^{M-1} \textmd{exp}\bigg(-sg_i w^{-\alpha}_i \bigg)\bigg]\\\nonumber
&\stackrel{(b)}{=}\mathbb{E}_{W}\bigg[\prod_{i\in\Phi_{st}} \bigg(\frac{m_i}{m_i+sw^{-\alpha}_i}\bigg)^{m_i}\prod_{j\in\Phi_{mo}} \bigg(\frac{m_j}{m_j+sw^{-\alpha}_j}\bigg)^{m_j}\bigg],
%&=\left[\mathbb{E}_{W_i}\left(\frac{m_i}{m_i+sw^{-\alpha}_i}\right)^{m_i}\right]^{M-1},
\end{align}
where $(a)$ follows the independence among $g_i$ and $w_i$, $(b)$ follows by taking the expectation over independent random variables $g_i$ for Nakagami-\emph{m} channels. Note that the number of UAVs in $\Phi_s$ and $\Phi_m$ are not independent, rather follow the binomial distribution. 
Finally, taking the expectation over the joint pdf of $f_{W_i}(w_i)$, (\ref{lt}) can be reached.

\section{Proof of Proposition \ref{tho1}}\label{appA}%
\begin{figure}[!t]
	\centering
	\includegraphics[width=3in]{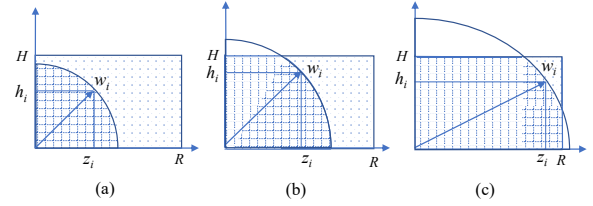}
	\caption{Illustration of integral regions in Proposition \ref{tho1} for cases: (a) $0\leq w_i<H$, (b) $H\leq w_i <R$, and (c) $R\leq w_i \leq \sqrt{R^2+H^2}$.}
	\label{intreg}
\end{figure}

With $W^2_i=h^2_i+Z^2_i$, the cdf of $W_i$ can be obtained as
\begin{align}\label{wym}
F_{W_i}(w_i)&=\textmd{Pr}[W_i<w_i]=\textmd{Pr}\big[Z_i<\sqrt{w^2_i-h^2_i}\big].
\end{align}
To evaluate (\ref{wym}), we need to consider three different cases (a), (b), and (c) as illustrated in Fig. \ref{intreg}. Therefore, we can write 
\begin{align}\label{wmr}
F^{st}_{W_i}(w_i)\!&=\!\left\{ \begin{array}{l}
\int_{0}^{w_i}\int_{0}^{\sqrt{w^2_i-h^2_i}}f_{Z_i}(z)f^{st}_{h_{i}}(x)dzdx,\\ 
\qquad\textmd{for } 0\leq w_i<H,\\
\int_{0}^{H}\int_{0}^{\sqrt{w^2_i-h^2_i}}f_{Z_i}(z)f^{st}_{h_{i}}(x)dzdx,\\ 
\qquad\textmd{for } H\leq w_i<R,\\
\int_{0}^{\sqrt{w^2_i-R^2}}\int_{0}^{R}f_{Z_i}(z)f^{st}_{h_{i}}(x)dzdx\\ 
+\!\int_{\sqrt{w^2_i-R^2}}^{H}\!\int_{0}^{\sqrt{w^2_i-h^2_i}}\!\!f_{Z_i}(z)f^{st}_{h_{i}}(x)dzdx\\
\qquad\textmd{for } R\leq w_i<\sqrt{R^2+H^2}.\\
%0, \textmd{ else}.
\end{array}\right.
\end{align}
In (\ref{wmr}), substituting the pdf of distance $Z_i$ as $f_{Z_i}(z)=\frac{2z}{R^2}$ for $0\leq z\leq R$ \cite{vv} along with the pdf $f^{st}_{h_{i}}(x)$, and evaluating the resulting integrals, one can derive (\ref{th101}).  
%\begin{align}\label{wmn}
%F_{W_i}(w_i)&=\left\{ \begin{array}{l}
%-\frac{4}{5}\frac{w^{5}_i}{R^2 H^3}+\frac{3}{2}\frac{w^{3}_i}{R^2 H^2},\textmd{ for }0\leq w_i<H,\\
%\frac{w^2_i}{R^2}-\frac{3}{10}\frac{H^2}{R^2},\textmd{ for }H\leq w_i<R,\\
%\frac{w^2_i}{R^2}-\frac{3}{10}\frac{H^2}{R^2}-\frac{3}{2}\frac{(w^2_i-R^2)^2}{R^2 H^2}+\frac{4}{5}\frac{(w^2_i-R^2)^{\frac{5}{2}}}{R^2 H^3}\\
%\qquad\textmd{ for } R\leq w_i\leq\sqrt{R^2+H^2}.
%%0, \textmd{ else}.
%\end{array}\right.
%%&=\frac{-2}{h^3_{max}}(w^2_m-r^2_m)^{\frac{3}{2}}+\frac{3}{h^2_{ma{\tiny }x}}(w^2_m-r^2_m),\\\nonumber
%%&\qquad \textmd{for }r_m\leq w_m \leq \sqrt{r^2_m+h^2_{max}}
%\end{align}
%Now, taking the derivative of $F_{W_{i}}(w_i)$ in (\ref{wmn}) with respect to $w_i$, (\ref{th11}) can be attained.

\end{document}